\documentclass[aip,reprint,amsmath,amssymb,]{revtex4-1}

\usepackage{graphicx}
\usepackage{amssymb}
\usepackage{epstopdf}
\usepackage{dcolumn}
\usepackage{bm}
\usepackage{amsmath}
\usepackage{mathrsfs}
\usepackage{siunitx}
\usepackage{subfigure}
\usepackage{flafter}
\usepackage{caption}
\usepackage{tabularx}
\usepackage{float}
\usepackage{braket,mleftright}
\usepackage[font={small}]{caption}
\newcolumntype{C}[1]{>{\centering\let\newline\\\arraybackslash\hspace{0pt}}m{#1}}
\newcolumntype{P}[1]{>{\centering\arraybackslash}p{#1}}

\newcommand\Tstrut{\rule{0pt}{2.3ex}}
\begin{document}

\title{Strain control of the N\'eel vector in Mn-based antiferromagnets} 

\author{In Jun Park}
\email{ipark008@ucr.edu}
\affiliation{Department of Electrical and Computer Engineering, University of California, Riverside, CA 92521, USA}

\author{Taehwan Lee}
\affiliation{Department of Materials Science and Engineering, University of California, Los Angeles, CA 90095, USA}

\author{Protik Das}

\author{Bishwajit Debnath}
\affiliation{Department of Electrical and Computer Engineering, University of California, Riverside, CA 92521, USA}

\author{Greg P. Carman}
\affiliation{Department of Mechanical and Aerospace Engineering, University of California, Los Angeles, CA 90095, USA}

\author{Roger K. Lake}
\email{rlake@ece.ucr.edu}
\affiliation{Department of Electrical and Computer Engineering, University of California, Riverside, CA 92521, USA}

\date{\today}

\begin{abstract}
Control of the N\'eel vector in antiferromagnetic materials is one of the challenges preventing 
their use as active device components.
Several methods have been investigated such as exchange bias, electric current, and spin injection, 
but little is known about strain-mediated anisotropy.
This study of the antiferromagnetic {\it L}1$_0$-type MnX alloys MnIr, MnRh, MnNi, MnPd, and MnPt
shows that a small amount of strain effectively 
rotates the direction of the N\' eel vector by 90$^{\circ}$ for all of the materials. 
For MnIr, MnRh, MnNi, and MnPd, the N\'eel vector rotates within the basal plane. 
For MnPt, the N\'eel vector rotates from out-of-plane to in-plane under tensile strain. 
The effectiveness of strain control is quantified by a metric of efficiency
and by direct calculation of the magnetostriction coefficients.
The values of the magnetostriction coefficients are comparable with those from ferromagnetic materials.
These results indicate that strain is a mechanism that can be exploited for control 
of the N\'eel vectors in this family of antiferromagnets.
\end{abstract}

\maketitle 

There has been a rapidly increasing interest in the use of
antiferromagnetic (AFM) materials for use as active device elements
\cite{2018_Tserkovnyak_RMP,2017_AFM_spintronics_Jungwirth_PSSR,AFM_spintronics_Jungwirth_NNano16}.
AFMs are insensitive to parasitic electromagnetic and magnetic interference.
The dipolar coupling is minimal, since there is no net magnetic moment.
Their lack of macroscopic magnetic fields allows AFM devices and interconnects to be 
highly scaled with reduced cross talk and insensitivity to geometrical anisotropy effects.
AFM resonant frequencies and magnon velocities are several orders of magnitude higher than those
in ferromagnetic materials, and these velocities correlate with similarly higher switching speeds
\cite{gomonay2014spintronics,AFM_spintronics_Jungwirth_NNano16,KWang_ULowSwitchingAFM_APL16}.
AFM metals and insulators are plentiful, and many have N\'{e}el temperatures well above room temperature,
a requirement for compatibility with on-chip temperatures in current Si integrated circuits.

\begin{figure}[tb]
\includegraphics[width=1.0\linewidth]{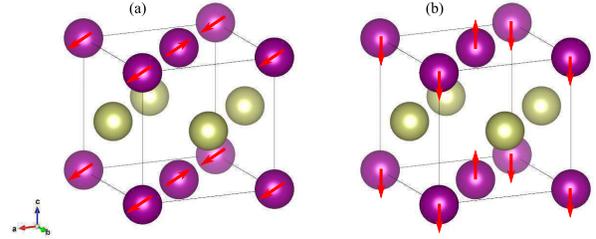}
\caption{\label{fig:structure} 
Antiferromagnetic {\it L}1$_0$-type Mn alloy structures. 
Mn atoms are the purple spheres with the spin vectors, and the gold spheres indicate the Ir, Rh, Ni, Pd, or Pt atoms. 
(a) In-plane equilibrium spin texture of MnIr, MnRh, MnNi, and MnPd.
(b) Out-of-plane equilibrium spin texture of MnPt. 
}
\end{figure}

The high N\'eel temperatures of the Mn-based equiatomic alloys such as MnIr, MnRh, MnNi, MnPd, and MnPt 
make them suitable candidates for on-chip applications \cite{2018_Tserkovnyak_RMP}. 
Extensive research has been conducted on the electronic \cite{sakuma1998electronic,umetsu2002electrical,umetsu2004pseudogap,umetsu2006electrical,umetsu2007electronic}, 
magnetic \cite{pal1968magnetic,sakuma1998electronic,umetsu2006electrical,umetsu2007electronic}, 
and elastic properties \cite{wang2013first,wang2013structural} of these materials.
The spins on the Mn 
atoms are antiferromagnetically coupled with each other in the basal plane, 
and each plane is coupled ferromagnetically as shown in Fig. \ref{fig:structure}.

The positive attributes of speed, scaling, and robustness to stray fields are accompanied by the challenges
of manipulating and detecting the antiferromagnetic states.
There are several methods to control the magnetic
properties of AFM materials such as with exchange bias \cite{2018_Tserkovnyak_RMP}, the use of electric current 
\cite{wadley2016electrical}, and strain induced by a 
piezoelectric material \cite{barra2018voltage,yan2019piezoelectric}. 
The recent experimental demonstration of strain control of the N\'eel vector in MnPt \cite{yan2019piezoelectric},
provides timely motivation for a theoretical study of strain-meditated magnetic anisotropy in the MnX AFM materials.
Density functional theory (DFT) is used to analyze the effect of strain on the magnetic anisotropy.
The effectiveness of strain control is quantified by a metric of efficiency and by calculation of the
magnetostriction coefficients.
%

\begin{table}[thb]
\centering
\caption{Calculated structure and local magnetic moment of the $L1_0$-type MnX alloys in the absence of strain.}
\label{table:str}
\begin{tabular}{c c c c c}
\hline\hline
\Tstrut
&  
a ({\AA}) & 
b ({\AA}) &
c ({\AA}) &
$\mu_{Mn}$ ({$\mu_{B}$})\\
\hline
MnIr&3.84&3.84&3.64&2.8 \\
MnRh&3.85&3.85&3.62&3.1 \\
MnNi&3.62&3.62&3.58&3.2 \\
MnPd&3.99&3.99&3.69&3.8 \\
MnPt&3.98&3.98&3.71&3.7 \\
\hline\hline
\end{tabular}
\label{tab1}
\end{table}
%
First principles calculations are performed as implemented in the Vienna Ab initio Simulation Package (VASP) \cite{kresse1993ab} to 
investigate the effect of strain on the magnetic anisotropy of {\it L}1$_0$-ordered bulk MnIr, MnRh, MnNi, MnPd, and MnPt. 
Projector augmented-wave (PAW) potentials \cite{blochl1994projector} and the generalized gradient approximation (GGA) 
parameterized by Perdew-Burke-Ernzerhof (PEB) were employed \cite{perdew1996generalized}. 
Depending on the materials, different cut-off 
energies (typically ranging from 420 eV to 450 eV) and k-points grids were used in order to ensure the total energy converged within 
10$^{-7}$ eV per unit cell. 
The initial equilibrium structure consists of a tetragonal unit cell where the 
fractional coordinates of Mn atoms are (0, 0, 0) and (0.5, 0.5, 0), 
and those of the X atoms are (0.5, 0, 0.5) and (0, 0.5, 0.5). 
Compressive or tensile stress along the $a$ axis is applied to each structure, 
and the structure is fully relaxed along the $b$ and $c$ axes (biaxially) 
until all forces on each atom are less than 10$^{-4}$ eV\AA$^{-1}$.  
The relaxed lattice constants for each applied strain are shown in supplementary Fig. S1.
The strain is defined as, ${\rm strain} = (a - a_0) / a_0 \times 100 \% $, 
where $a$ and $a_0$ are the lattice constants with and without strain, respectively. 
With the relaxed structure, 
the spin-polarized self-consistent calculation is performed to obtain the charge density. 
Finally, the magnetic anisotropy energies are determined by calculating the total 
energies for different N\'eel vector directions including spin orbit coupling.
Table \ref{tab1} shows the lattice constants and the magnetic moments of the Mn site in MnX without strain. 
All of the values are very close to those from previous results \cite{wang2013first,wang2013structural}. 
The local magnetic moments of the X site are zero for all materials. 
%

\begin{figure}
\includegraphics[width=.8\linewidth]{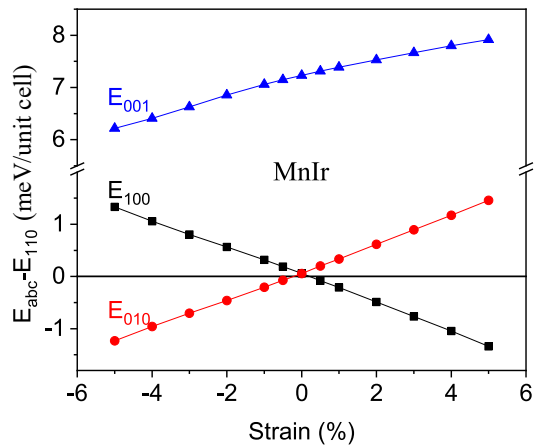}
\caption{\label{fig:Ir} 
MnIr energy differences $E_{abc} - E_{110}$ for the 3 different orientations of the
N\'eel vector as indicated by the labels.
}
\end{figure}
\begin{figure}
\includegraphics[width=.8\linewidth]{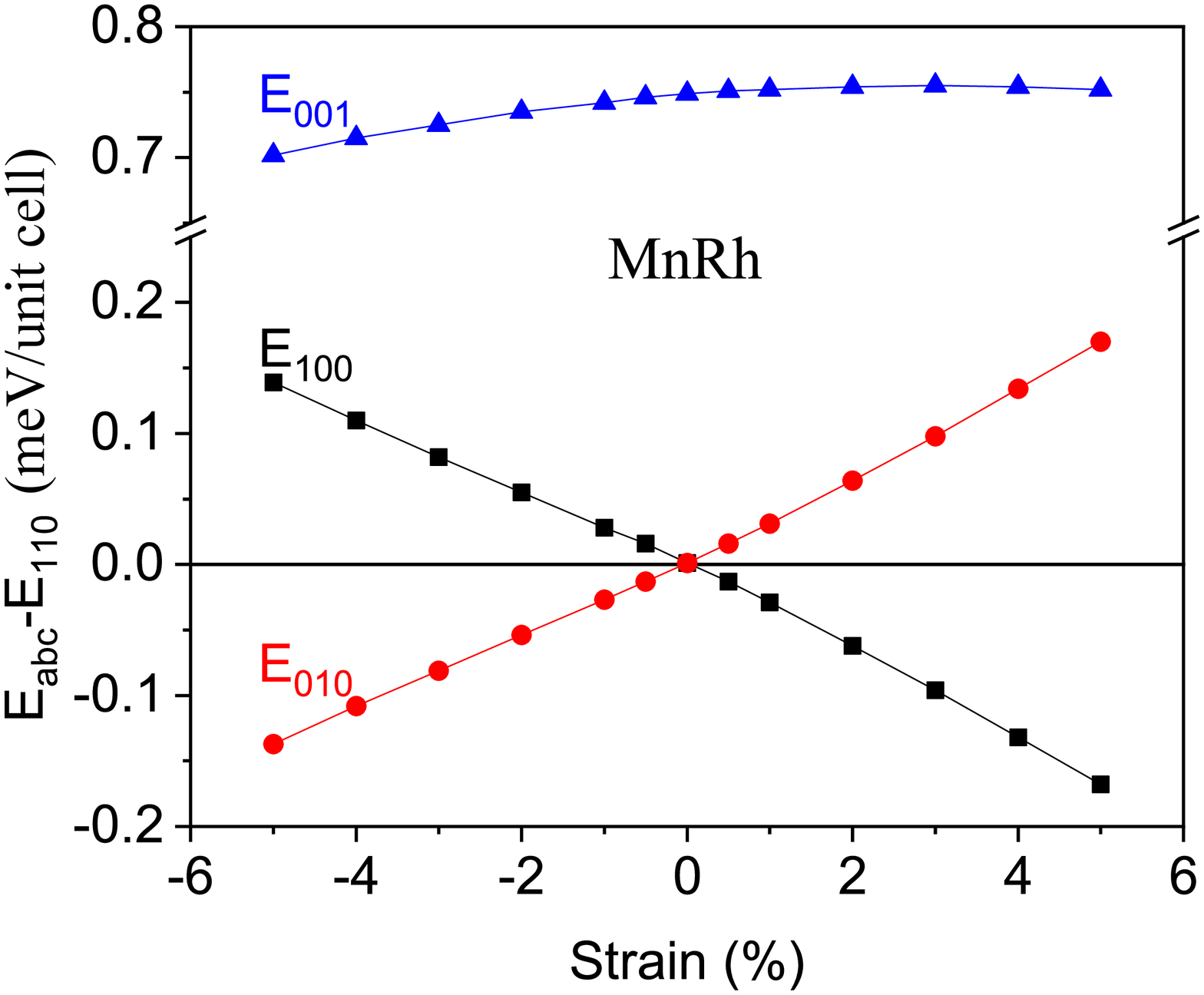}
\caption{\label{fig:Rh} 
MnRh energy differences $E_{abc} - E_{110}$ for the 3 different orientations of the
N\'eel vector as indicated by the labels.
}
\end{figure}
\begin{figure}
\includegraphics[width=.8\linewidth]{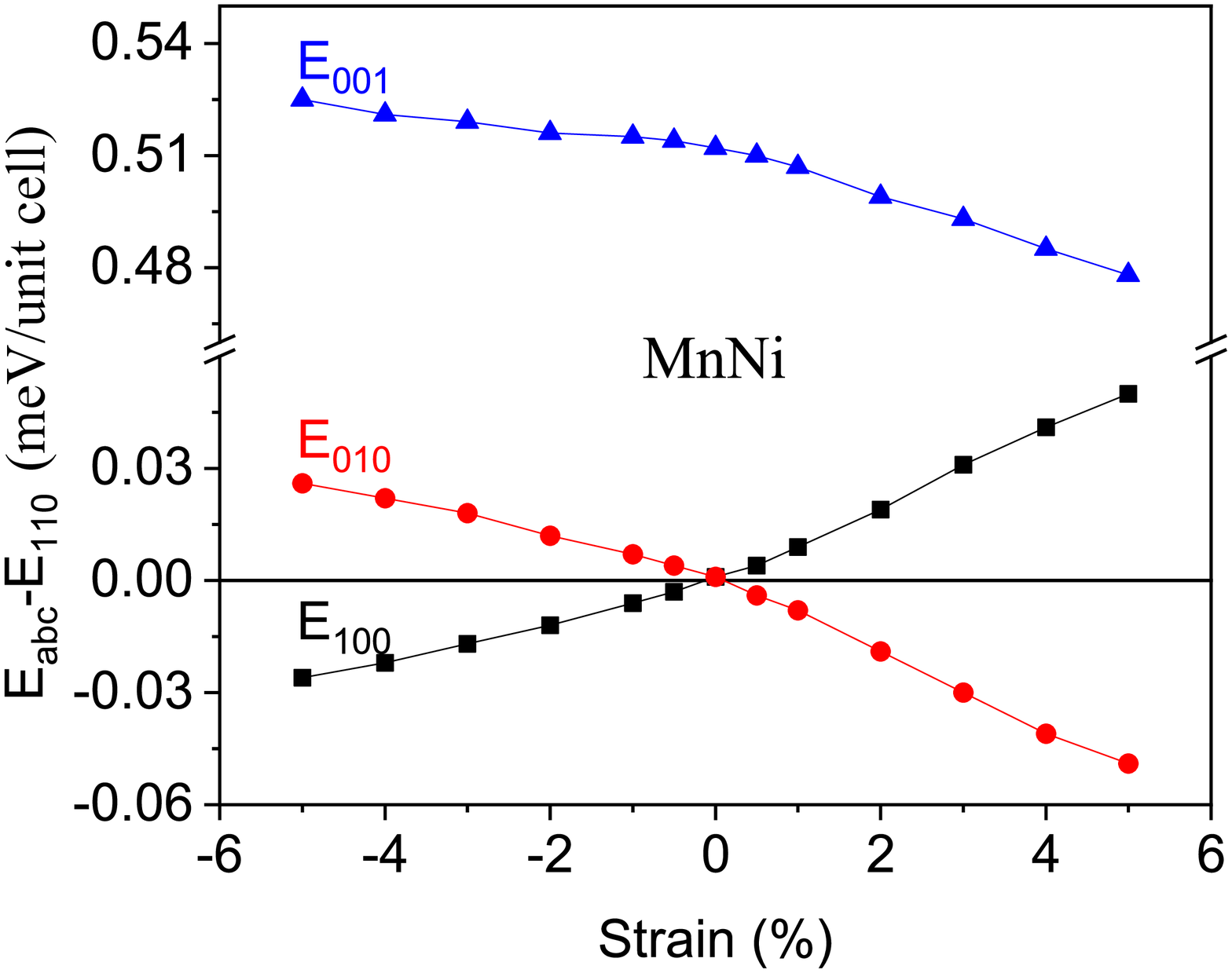}
\caption{\label{fig:Ni} 
MnNi energy differences $E_{abc} - E_{110}$ for the 3 different orientations of the
N\'eel vector as indicated by the labels.
}
\end{figure}
\begin{figure}
\includegraphics[width=.8\linewidth]{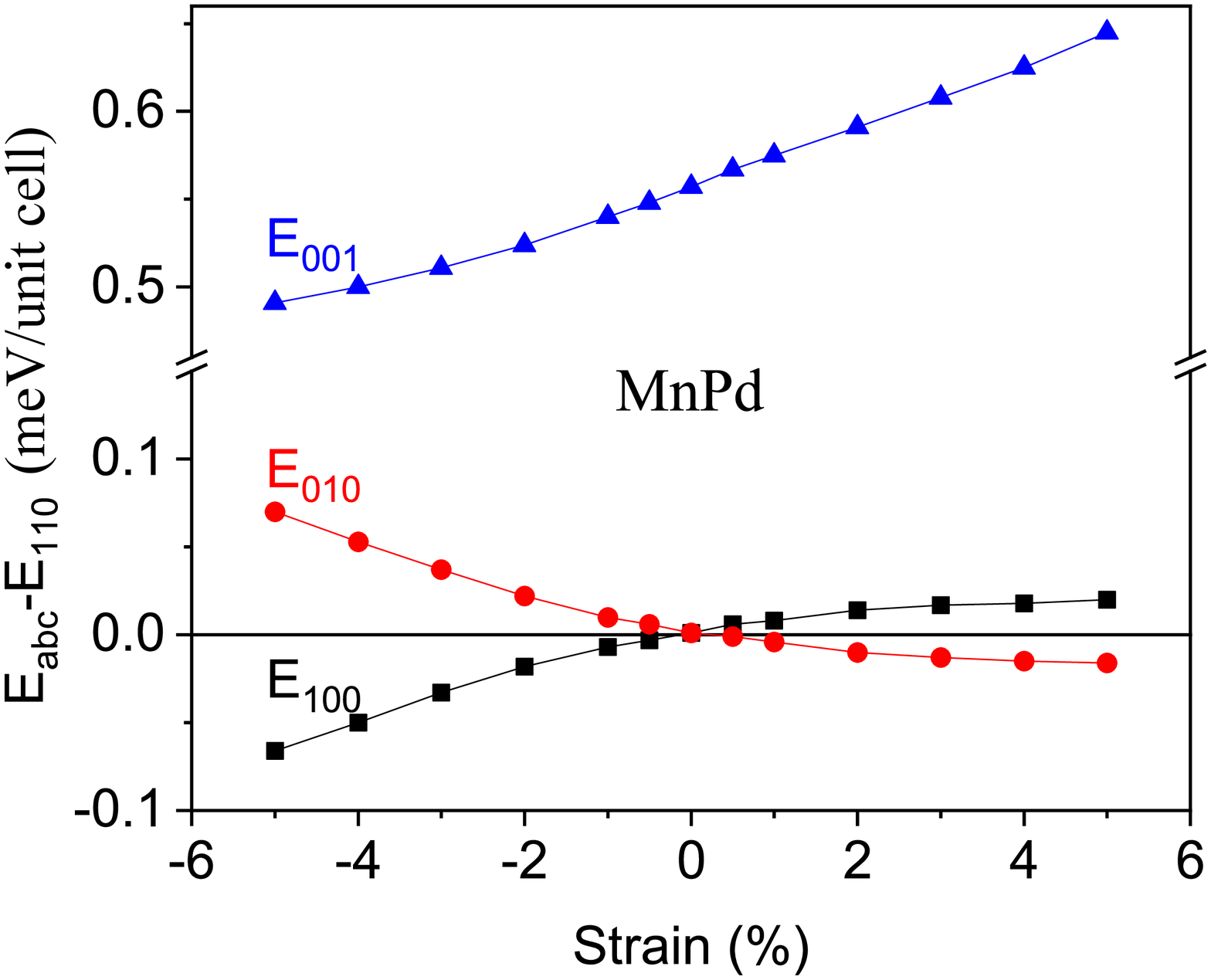}
\caption{\label{fig:Pd} 
MnPd energy differences $E_{abc} - E_{110}$ for the 3 different orientations of the
N\'eel vector as indicated by the labels.
}
\end{figure}
\begin{figure}
\includegraphics[width=.8\linewidth]{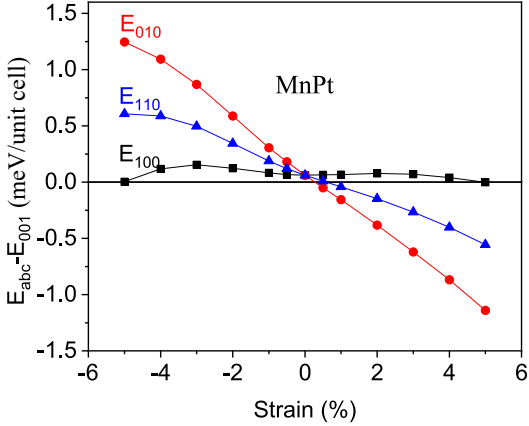}
\caption{\label{fig:Pt} 
MnPt energy differences $E_{abc} - E_{110}$ for the 3 different orientations of the
N\'eel vector as indicated by the labels.
}
\end{figure}

Figures \ref{fig:Ir}--\ref{fig:Pt} 
show the differences in the total energies as a function of the strain for
MnIr, MnRh, MnNi, MnPd, and MnPt, respectively, 
where $E_{abc}$ is the ground state energy with the N\'eel vector along the $[abc]$ direction. 
The reference energy levels from each figure, which are indicated by the solid black lines, are $E_{001}$ for 
MnPt and $E_{110}$ for the other materials. 
The reference energies are the lowest energy state, which means MnIr, MnRh, MnNi, and 
MnPd have in-plane anisotropy and MnPt has out-of-plane anisotropy without strain. 
This is consistent with experimental results \cite{pal1968magnetic}. 
To show the energy differences more clearly as the strain changes, the reference level is taken at each value of the 
applied strain. 
At zero strain, there is no energy difference between $E_{100}$ and $E_{010}$ because of the symmetry of all of the materials.


Figures \ref{fig:Ir}--\ref{fig:Pd} show that sweeping the strain from negative (compressive) 
to positive (tensile) causes
a 90$^{\circ}$ rotation of the N\'eel vector in the $ab$-plane for the four materials
MnIr, MnRh, MnNi, and MnPd.
However, the alignment of the N\'eel vector with compressive or tensile strain depends on the specific material. 
MnIr and MnRh behave like magnets with a positive magnetostriction coefficient, 
since tensile strain along [100] causes the N\'eel vector to align in the [100] direction. 
On the other hand, MnNi and MnPd behave like magnets with a negative magnetostriction coefficient,
since tensile strain along [100] causes the N\'eel vector to align in the [010] direction \cite{biswas2014energy}. 

MnPt is unique among the 5 materials.
In equilibrium, in the absence of strain, the N\'eel vector has perpendicular anisotropy.
Under compressive (negative) strain along the $[100]$ axis, the N\'eel vector remains out-of-plane. 
Under tensile strain along $[100]$, the N\'eel vector switches from out-of-plane $[001]$ to in-plane 
aligning in the $[010]$ direction.

\begin{figure}
\includegraphics[width=1.0\linewidth]{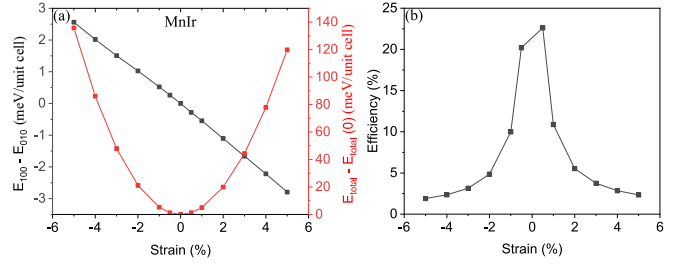}
\caption{\label{fig:Ireff} 
MnIr strain energies and efficiency versus strain.
(a) The energy difference between two different N\' eel vector orientations (black) as shown by the left axis, 
and the change in total energy (red) as shown by the right axis. 
(b) The efficiency as a function of the strain.
}
\end{figure}
\begin{figure}[tb]
\includegraphics[width=1.0\linewidth]{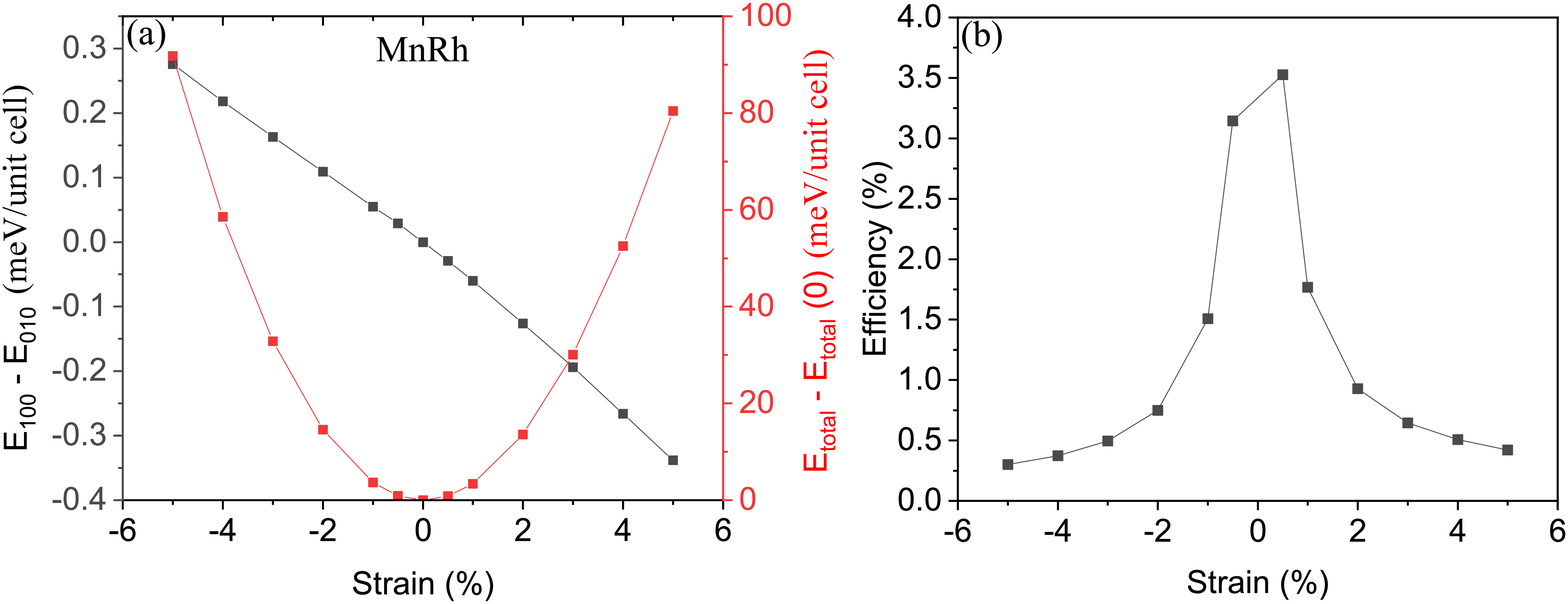}
\caption{\label{fig:Rheff} 
MnRh strain energies and efficiency versus strain.
(a) The energy difference between two different N\'eel vector orientations (black) as shown by the left axis, 
and the change in total energy (red) as shown by the right axis. 
(b) The efficiency as a function of the strain.
}
\end{figure}
\begin{figure}[tb]
\includegraphics[width=1.0\linewidth]{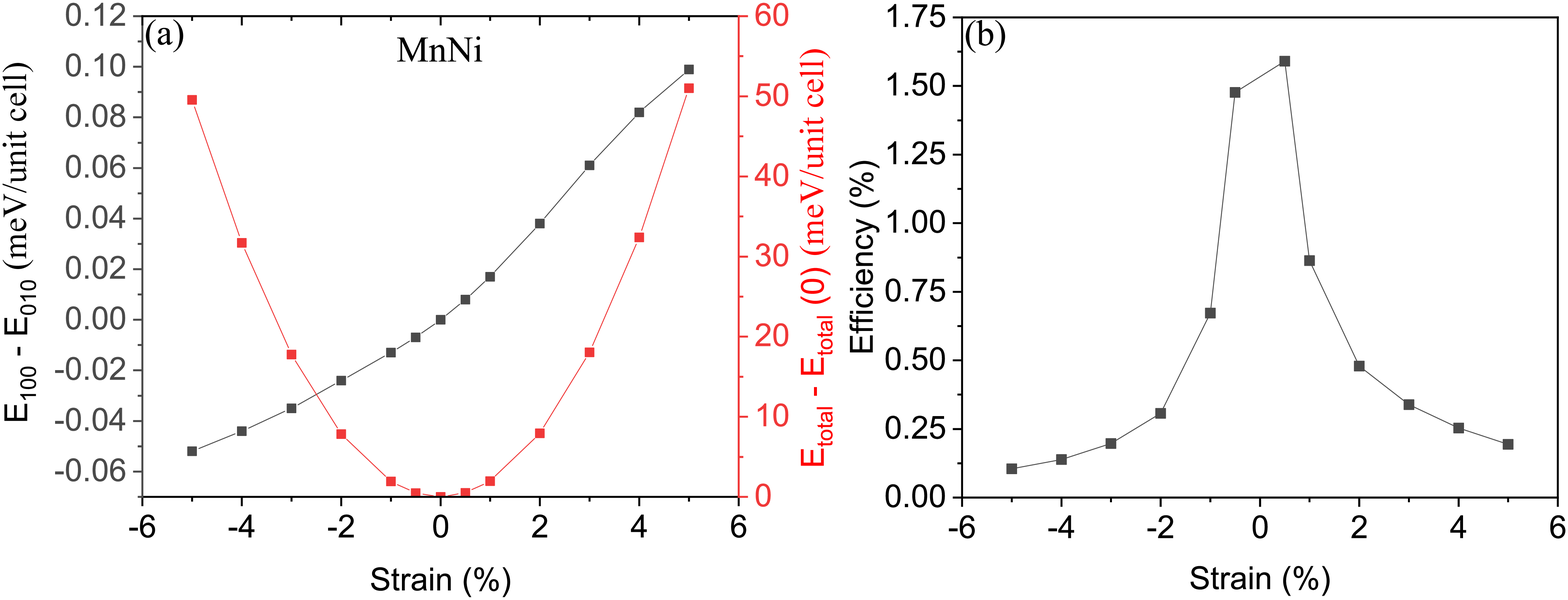}
\caption{\label{fig:Nieff} 
MnNi strain energies and efficiency versus strain.
(a) The energy difference between two different N\'eel vector orientations (black) as shown by the left axis, 
and the change in total energy (red) as shown by the right axis. 
(b) The efficiency as a function of the strain.
}
\end{figure}
\begin{figure}[tb]
\includegraphics[width=1.0\linewidth]{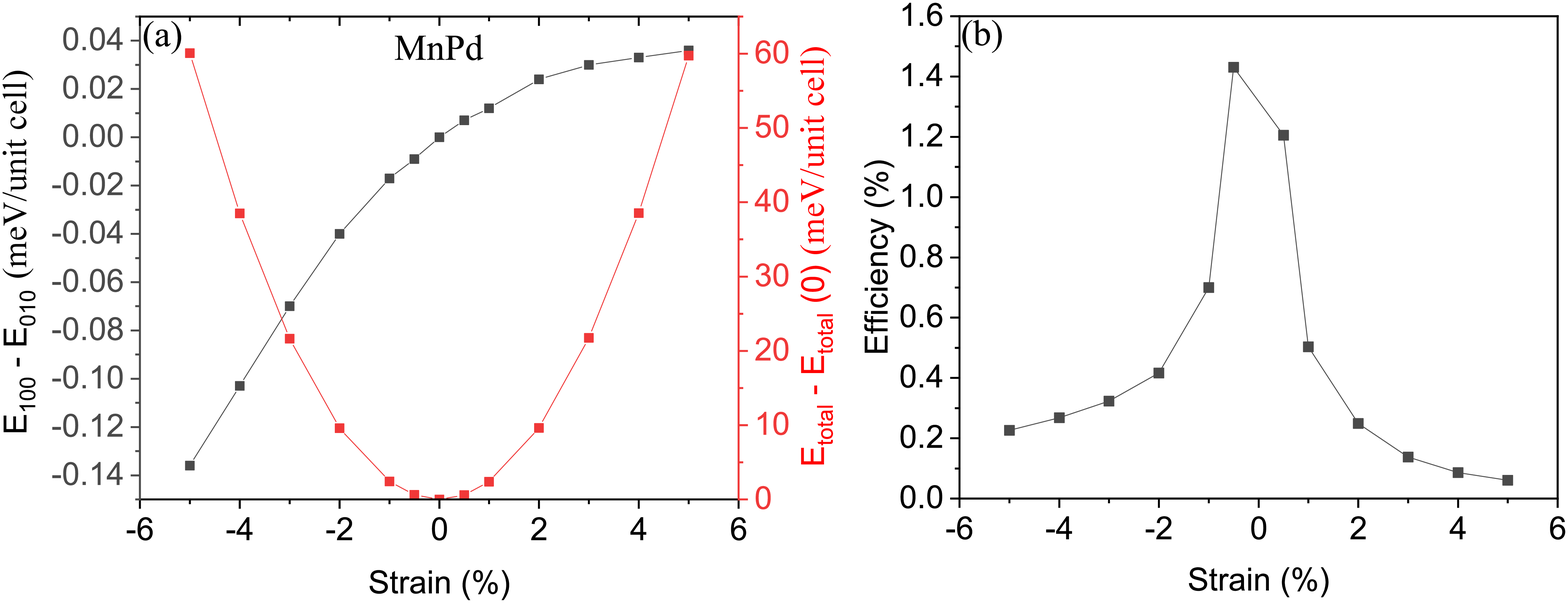}
\caption{\label{fig:Pdeff} 
MnPd strain energies and efficiency versus strain.
(a) The energy difference between two different N\'eel vector orientations (black) as shown by the left axis, 
and the change in total energy (red) as shown by the right axis. 
(b) The efficiency as a function of the strain.
}
\end{figure}
\begin{figure}[tb]
\includegraphics[width=1.0\linewidth]{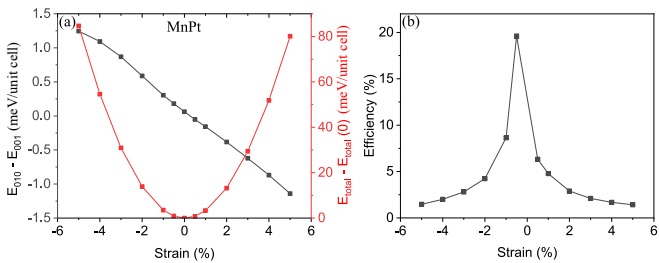}
\caption{\label{fig:Pteff} 
MnPt strain energies and efficiency versus strain.
(a) The energy difference between two different N\'eel vector orientations (black) as shown by the left axis, 
and the change in total energy (red) as shown by the right axis. 
(b) The efficiency as a function of the strain.
}
\end{figure}

For applications, it is useful to quantify the efficiency with which strain rotates the N\'eel vector
and to determine the magnetostriction coefficient from the ab initio calculations.
The internal efficiency is defined as
\begin{equation}
{\rm Efficiency} (\%) = \left| \frac{E_{abc} - E_{a'b'c'}}{E_{total} - E_{total}(0)} \right| \times 100 ,
\label{eq:efficiency}
\end{equation}
where the total energies $E_{abc}$ and $E_{a'b'c'}$ are defined in the same way as above,
i.e. the total energies in the presence of strain with the N\'eel vector oriented along $[abc]$ or $[a'b'c']$, 
respectively. 
The denominator in the Eq. (\ref{eq:efficiency}) is the total energy change induced by the strain. 
For MnIr, MnRh, MnNi, and MnPd, $E_{abc}$ and $E_{a'b'c'}$ are $E_{100}$ and $E_{010}$, respectively.
For MnPt, $E_{abc}$ and $E_{a'b'c'}$ are $E_{010}$ and $E_{001}$, respectively.
The numerator and denominator of Eq. (\ref{eq:efficiency}) are plotted
as a function of strain in Figs. \ref{fig:Ireff}-\ref{fig:Pteff}(a),
and the resulting efficiencies are plotted as a function of strain in Figs. \ref{fig:Ireff}-\ref{fig:Pteff}(b).
The changes in the total energies, shown as red curves in Figs. \ref{fig:Ireff}-\ref{fig:Pteff}(a),
are parabolic so that they can be considered as the strain energy proportional to the square of the applied strain. 
On the other 
hand, the differences between two energies (the black curves in Figs. \ref{fig:Ireff}-\ref{fig:Pteff}(a)) 
are approximately linear under small strain ($< 1\%$).
Therefore, the efficiency decreases sharply as the amount of strain increases. 
At $0.5\%$ strain, the highest efficiency for $90^\circ$ in-plane rotation of the N\'eel vector is $20\%$ for MnIr.
For MnRh, MnNi, and MnPd, the efficiencies are smaller and equal to $3.5\%$, $1.5\%$, and $1.4\%$, respectively.
To rotate the N\'eel vector from out-of-plane to in-plane in MnPt, a positive, tensile strain must be applied.
The efficiency of this process at $+0.5\%$ strain is $6\%$. 

Using the data above, the magnetostriction coefficients ($\lambda_s$), 
which are widely used in ferromagnets, are calculated. 
The magnetostriction coefficient is defined as
\begin{equation}
\lambda_s (ppm) = \frac{2K_{me}}{3Y(\varepsilon_{bb} - \varepsilon_{aa})},
\label{eq:coefficient}
\end{equation}
where $Y$ and $(\varepsilon_{bb} - \varepsilon_{aa})$ are 
Young's modulus and strain, respectively \cite{bur2011strain}.
$K_{me}$ is the magnetoelastic anisotropy constant, 
which is defined as the difference of the magnetic anisotropy energies with and without strain,
and the magnetic anisotropy energy is defined as $E_{100} - E_{010}$.
Plots of $E_{100} - E_{010}$ as a function of strain are shown in Supplementary Fig. S2.
Young's moduli for all MnX alloys except MnIr were adopted from previous calculation results 
\cite{wang2013first,wang2013structural}, 
and the value for MnIr was determined as described in the Supplementary Information.
For simplicity, we disregard $\varepsilon_{bb}$ which represents a negligible change in the 
lattice constant along the $b$-axis caused by the applied strain along $a$.
The results for $\lambda_s$ are summarized in the Table \ref{tab2}. 
As expected, MnIr and MnRh 
have positive values of $\lambda_s$, and MnNi, MnPd, and MnPt have negative values. 
Also, the magnitudes of the magnetostriction coefficients follow the magnitudes of the efficiencies.
The magnetostriction coefficients of the MnX alloys are 
comparable with the ones from ferromagnets 
\cite{clark2000magnetostrictive,panduranga2018polycrystalline,hall1959single,huang1995giant,fritsch2012first}, 
which suggests that strain can be used to control the magnetic anisotropy of these 
antiferromagnetic materials.
\begin{table}[H]
\centering
\caption{Calculated magnetrostriction coefficients of the $L1_0$-type MnX alloys.}
\label{table:coef}
\begin{tabular}{c c c c c c}
\hline\hline
  & MnIr & MnRh & MnNi & MnPd & MnPt\\
\hline
$\lambda_s$ (ppm)&241&43&-15&-17&-196\\
\hline\hline
\end{tabular}
\label{tab2}
\end{table}

%
In summary, the N\'eel vectors of MnIr, MnRh, MnNi, and MnPd 
can be rotated $90^\circ$ in the basal plane by applying in-plane strain.
MnIr and MnRh behave like magnets with positive magnetostriction coefficients,
since their N\'eel vectors align with tensile strain.
MnNi and MnRh behave like magnets with negative magnetostriction coefficients, since
their N\'eel vectors align with compressive strain.
The internal efficiency of this process is highest for MnIr 
and it is equal to $20\%$ at $0.5\%$ strain.
MnPt is unique among the 5 alloys in that its N\'eel vector aligns out-of-plane
along the [001] axis in equilibrium.
Applying a tensile strain along [100] rotates the N\'eel vector from out-of-plane [001] to in-plane [010].
The efficiency of this process at $0.5\%$ tensile strain is $6\%$.
Under compressive strain along [100], the N\'eel vector of MnPt remains out-of-plane [001]. 
The magnitudes of the calculated magnetostriction coefficients are comparable with those of ferromagnets,
and they follow the same trends as the calculated efficiencies.
For in-plane rotation of the N\'eel vector, MnIr has the highest magnetostriction coefficient of 241 ppm. 
The magnetostriction coefficient for out-of-plane rotation in MnPt is -196 ppm.
These results suggest that strain can be an effective mechanism to control the N\'eel vectors in this family of antiferromagnets.

This work was supported as part of Spins and Heat in Nanoscale Electronic Systems (SHINES) an 
Energy Frontier Research Center funded by the U.S. Department of Energy, Office of Science, 
Basic Energy Sciences under Award \#DE-SC0012670. 
This material is based upon work supported by or in part by the 
U.S. Army Research Laboratory and the U.S. Army Research 
Office under Grant No. W911NF-17-0364.
This work used the Extreme Science and Engineering Discovery
Environment (XSEDE) \cite{towns2014xsede}, which is supported by National
Science Foundation Grant No. ACI-1548562 and allocation
ID TG-DMR130081.

%

\end{document}


\title{Strain control of the N\'eel vector in Mn-based antiferromagnets: Supplementary Information} 

\author{In Jun Park}
\email{ipark008@ucr.edu}
\affiliation{Department of Electrical and Computer Engineering, University of California, Riverside, CA 92521, USA}

\author{Taehwan Lee}
\affiliation{Department of Materials Science and Engineering, University of California, Los Angeles, CA 90095, USA}

\author{Protik Das}

\author{Bishwajit Debnath}
\affiliation{Department of Electrical and Computer Engineering, University of California, Riverside, CA 92521, USA}

\author{Greg P. Carman}
\affiliation{Department of Mechanical and Aerospace Engineering, University of California, Los Angeles, CA 90095, USA}

\author{Roger K. Lake}
\email{rlake@ece.ucr.edu}
\affiliation{Department of Electrical and Computer Engineering, University of California, Riverside, CA 92521, USA}

\date{\today}


\maketitle 

\begin{figure}[H]
\centering
\includegraphics[width=.4\linewidth]{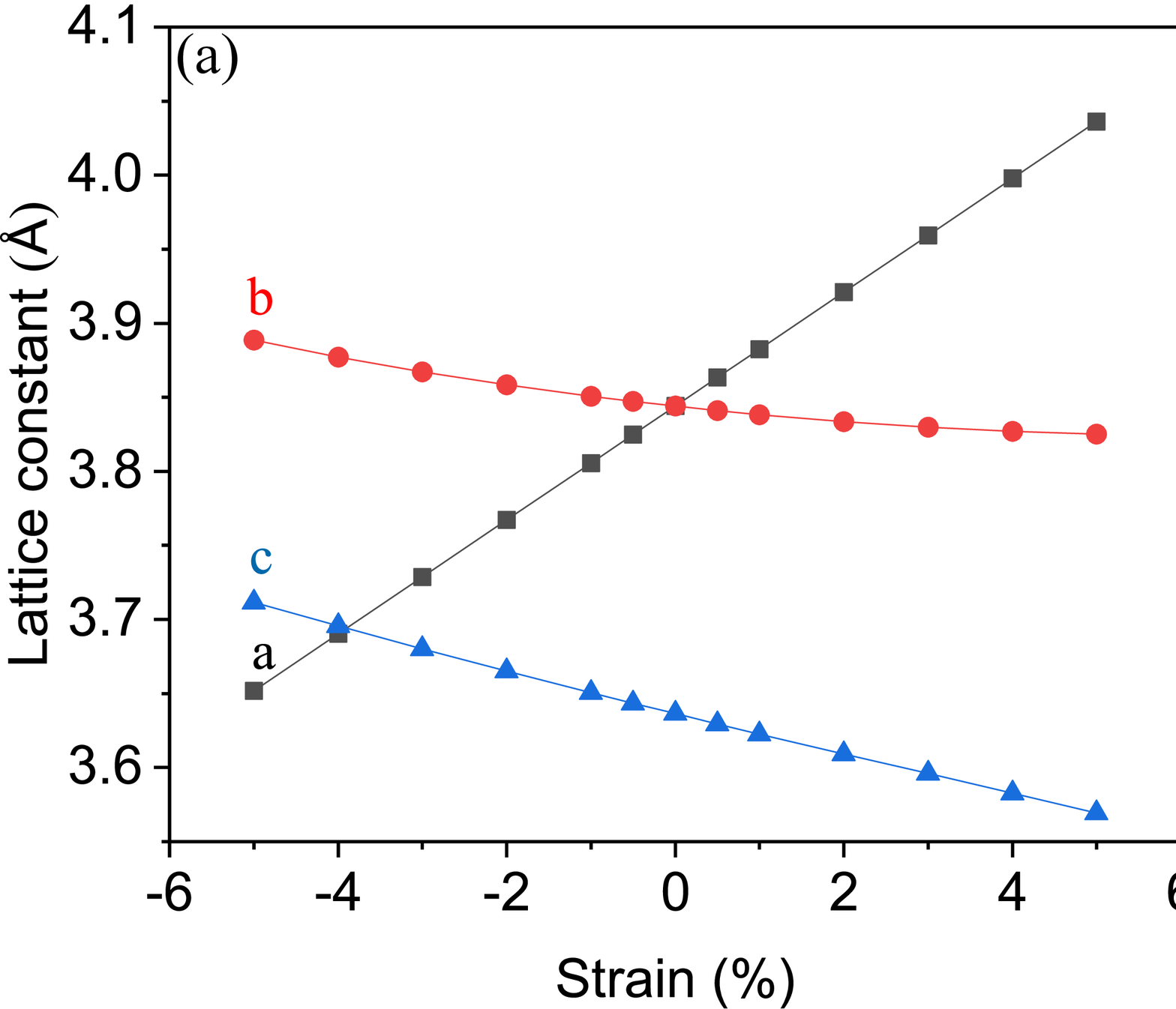}
\includegraphics[width=.4\linewidth]{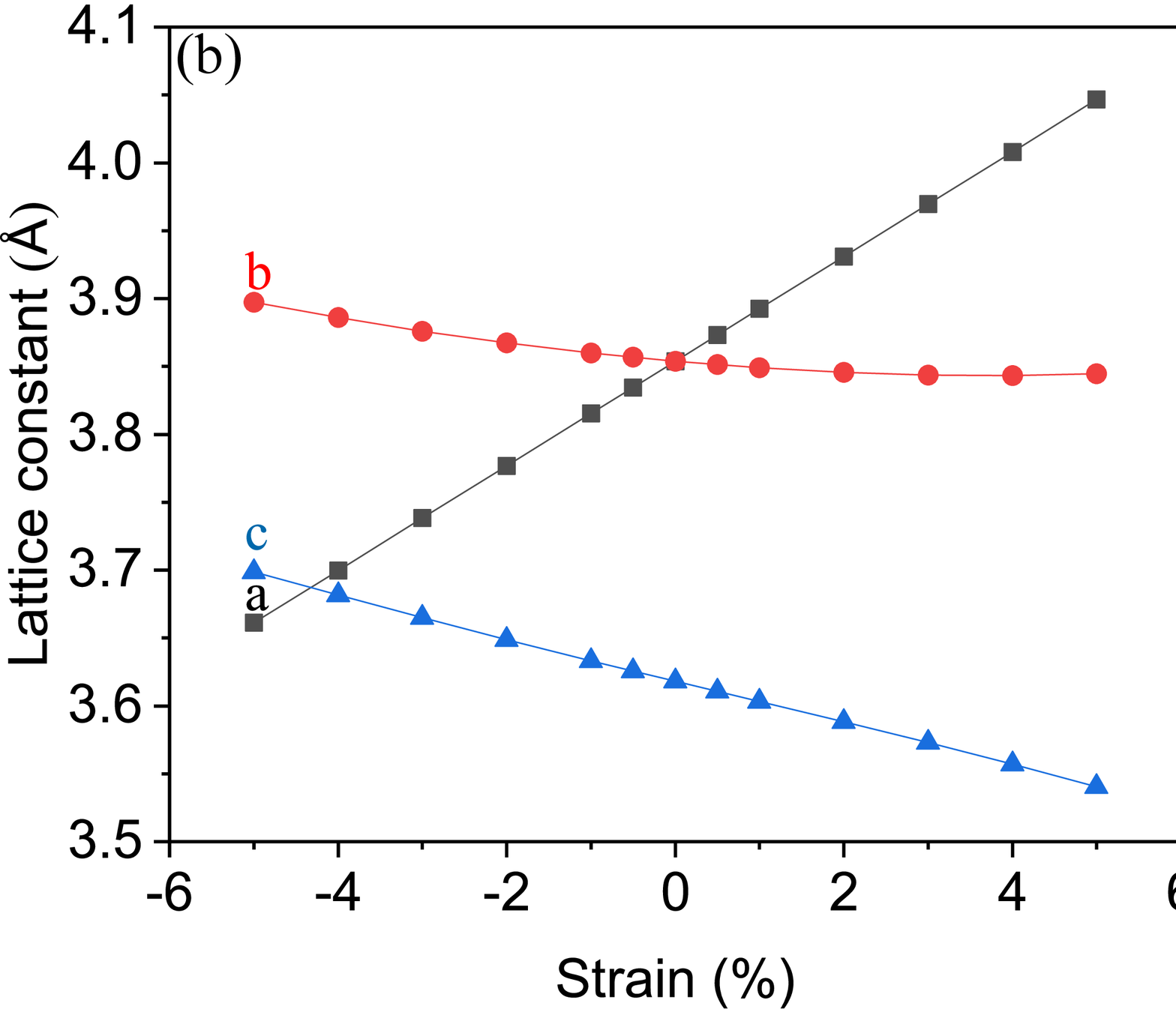}\\
\includegraphics[width=.4\linewidth]{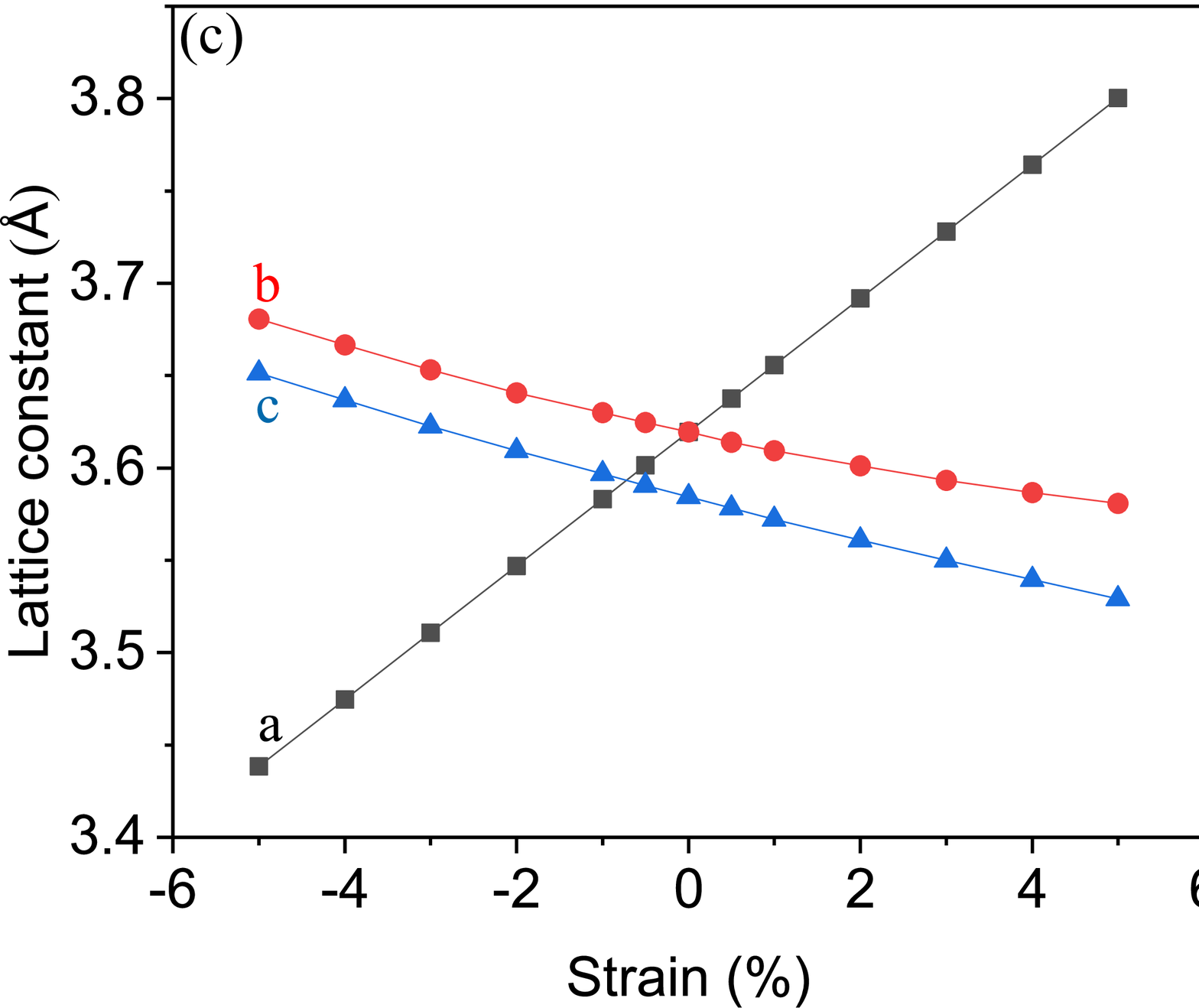}
\includegraphics[width=.4\linewidth]{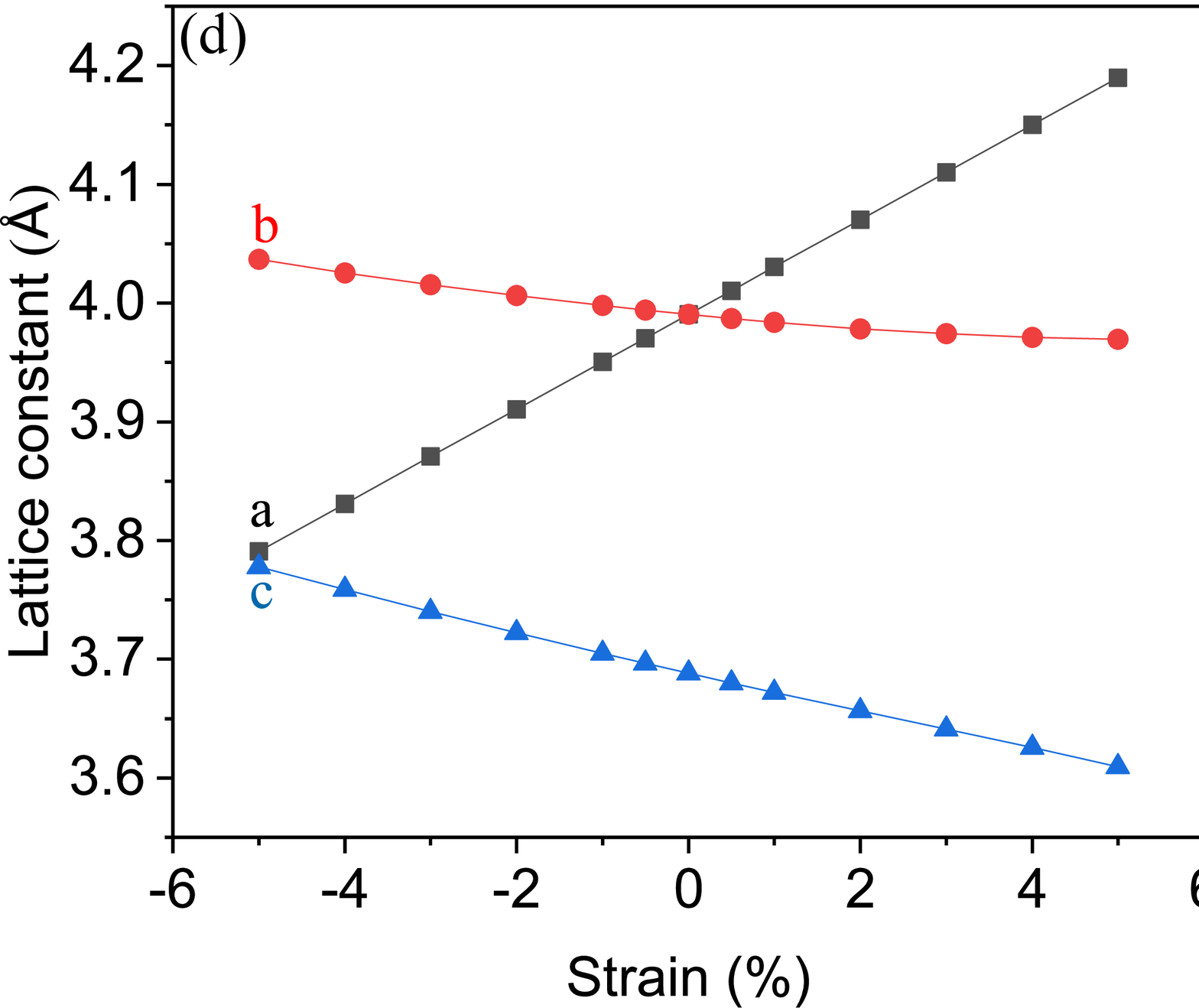}\\
\includegraphics[width=.4\linewidth]{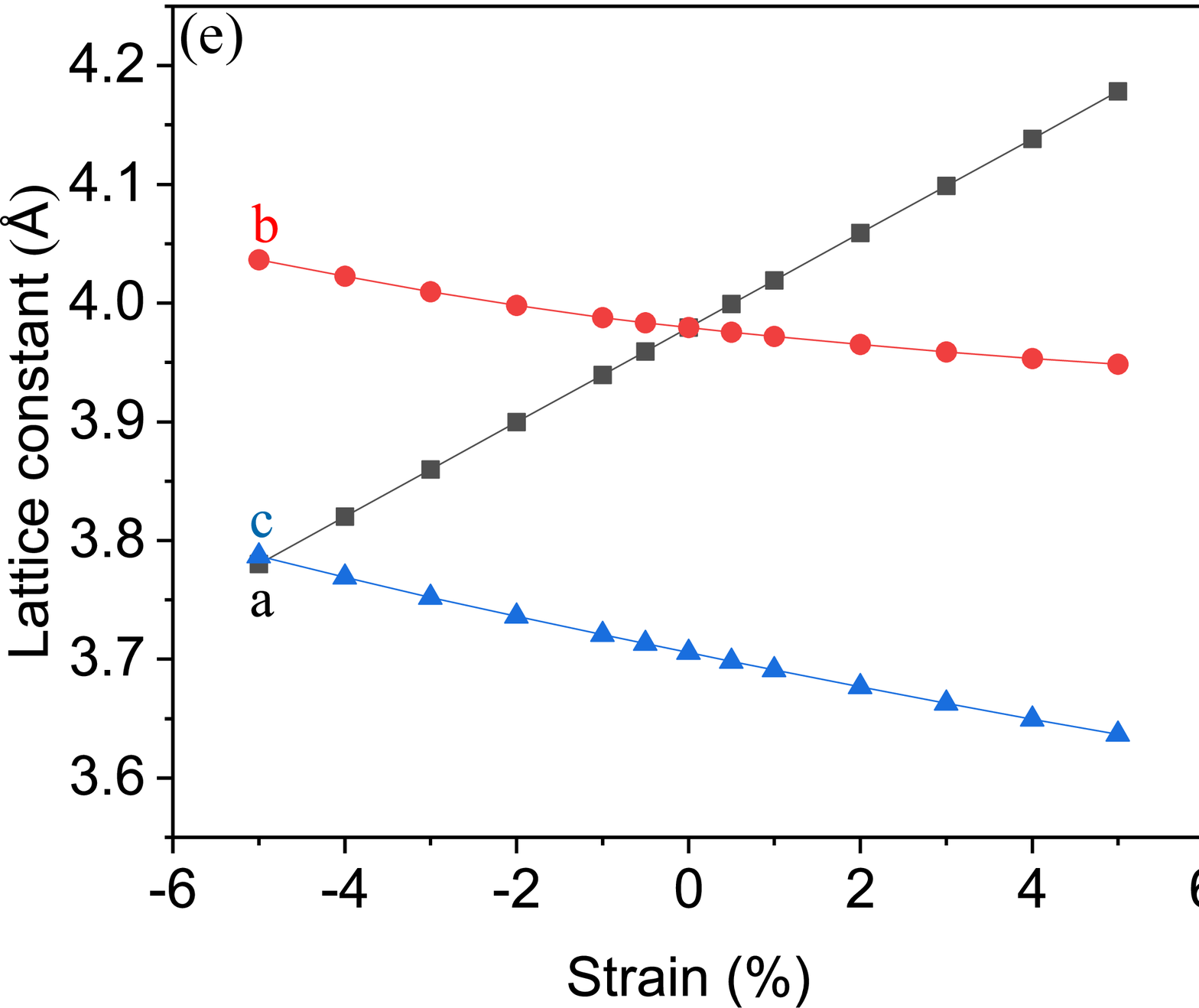}
\caption{\label{fig:constant} 
The lattice constants versus applied strain in (a) MnIr, (b) MnRh, (c) MnNi, (d) MnPd, and (e) MnPt.}
\end{figure}

\begin{figure}
\centering
\includegraphics[width=.4\linewidth]{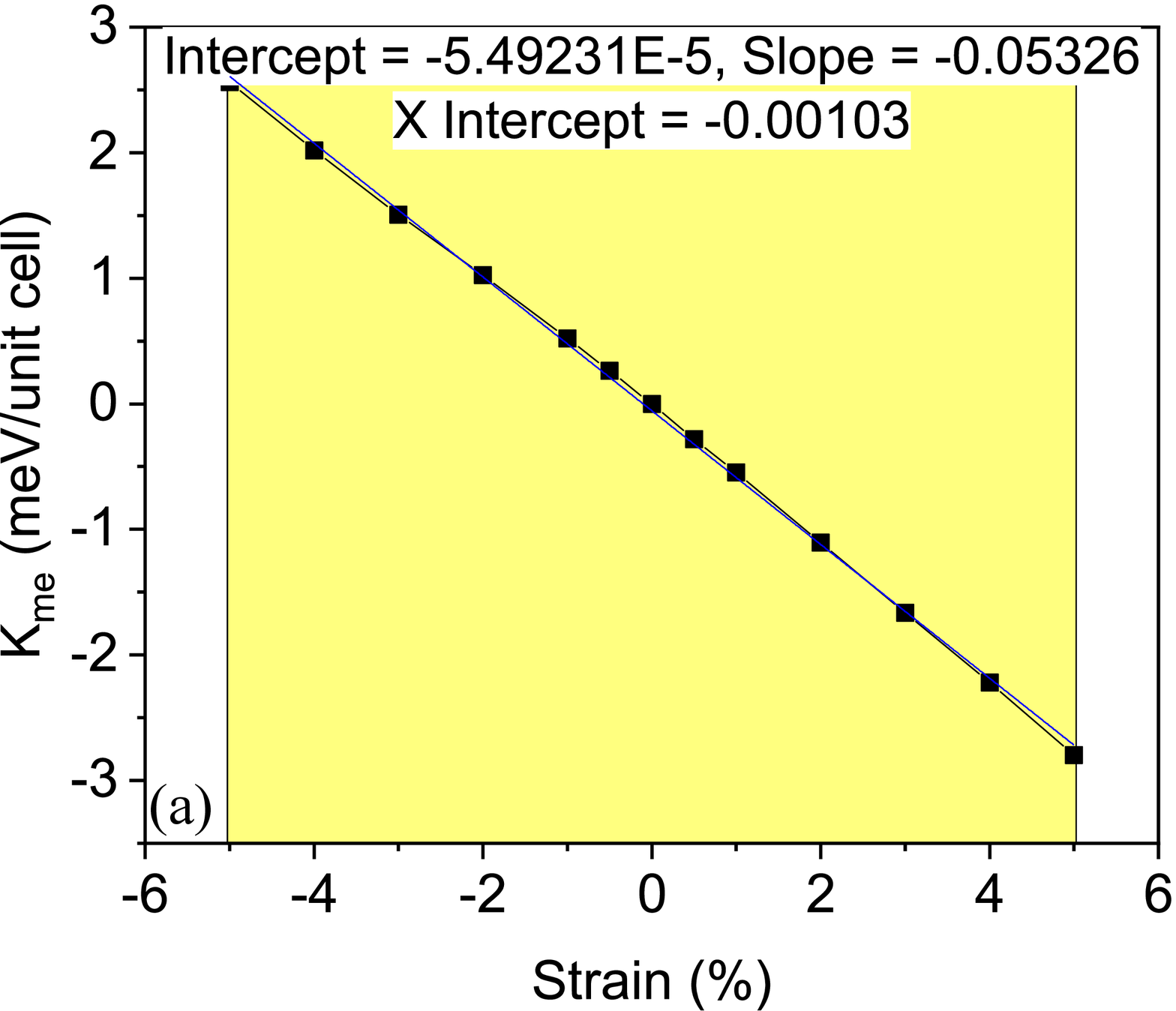}
\includegraphics[width=.4\linewidth]{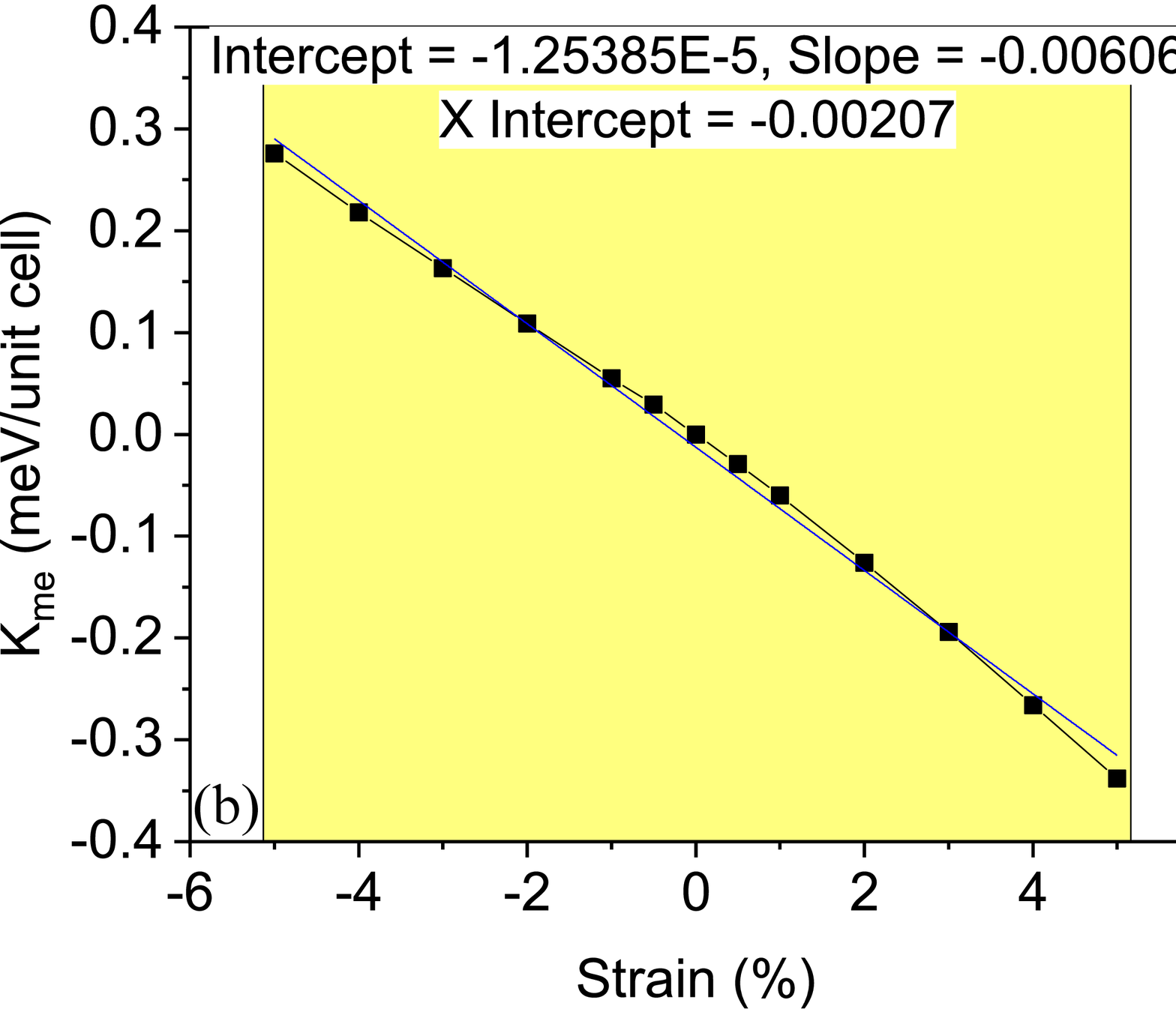}\\
\includegraphics[width=.4\linewidth]{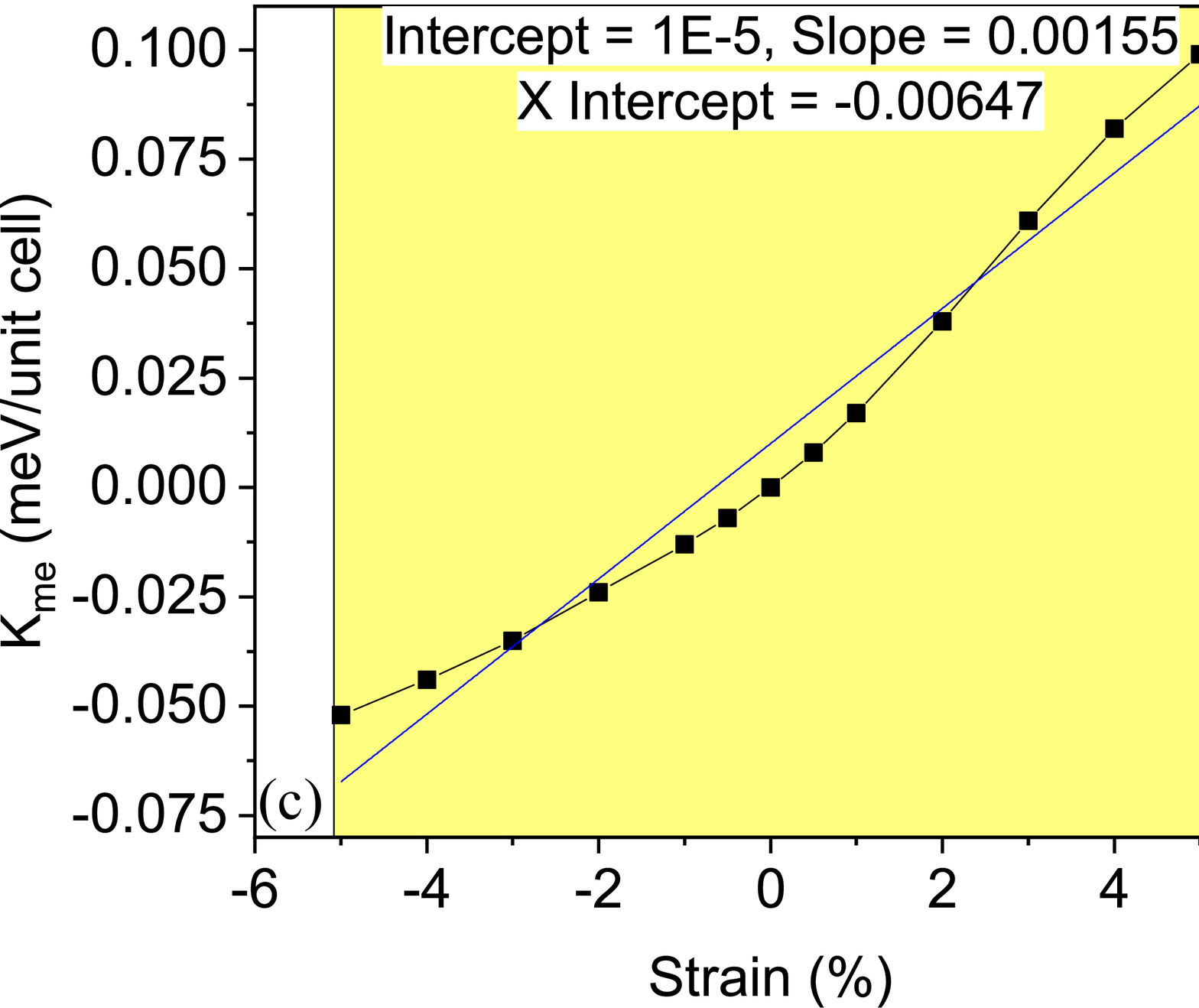}
\includegraphics[width=.4\linewidth]{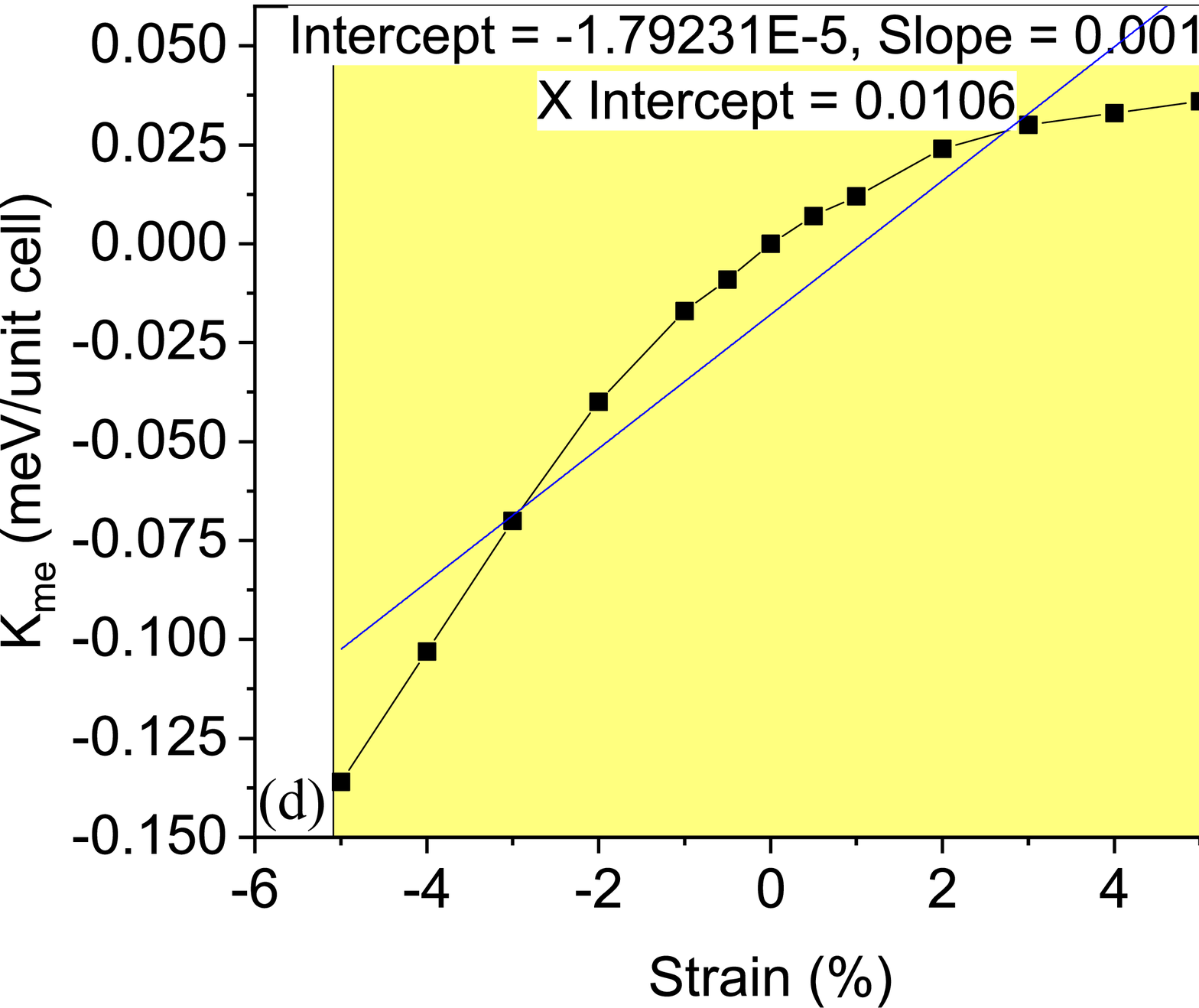}\\
\includegraphics[width=.4\linewidth]{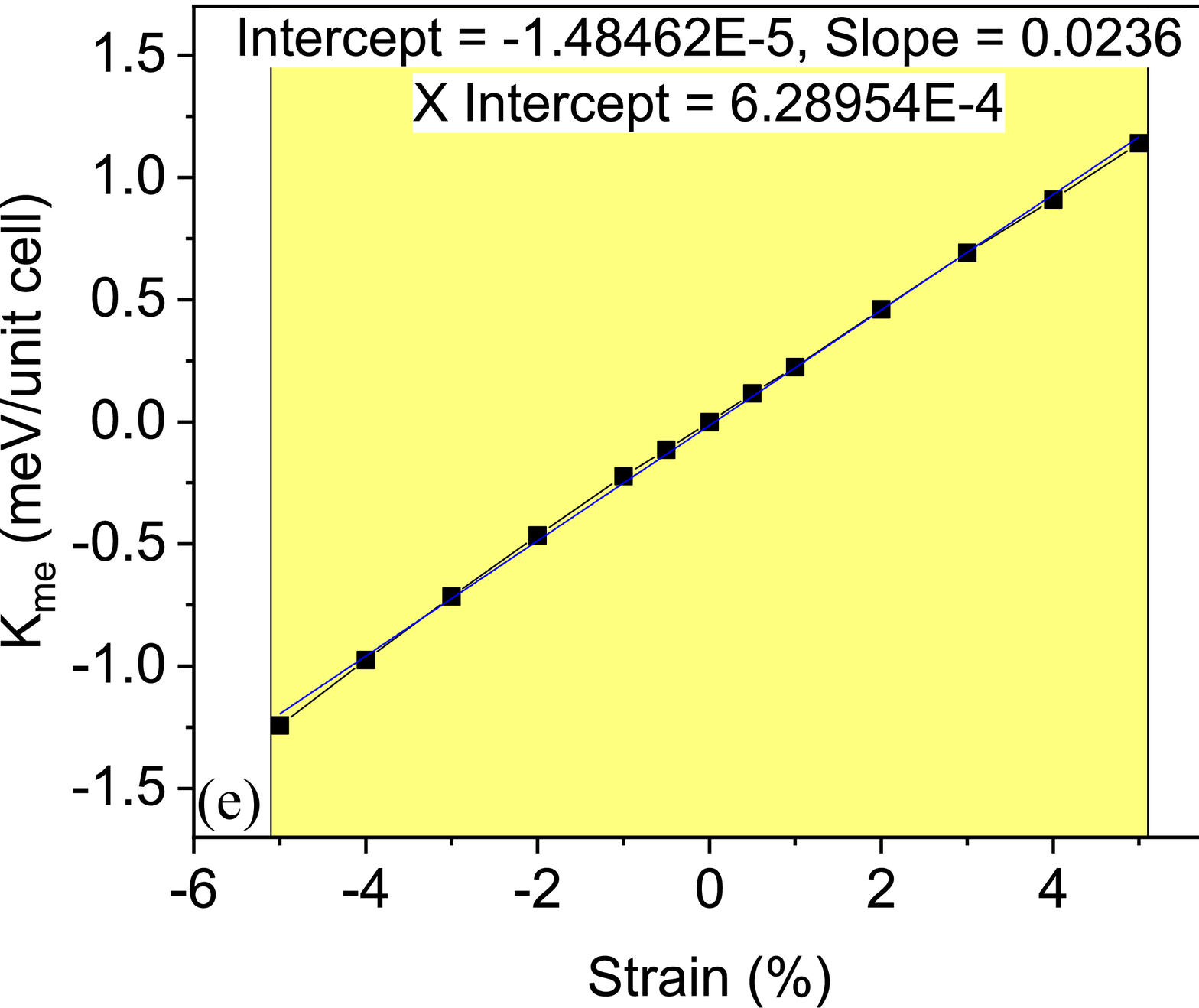}
\caption{\label{fig:coefficient} 
The magnetoelastic anisotropy constants versus applied strain in (a) MnIr, (b) MnRh, (c) MnNi, (d) MnPd, and (e) MnPt.}
\end{figure}

Figure \ref{fig:constant} shows the chage in the lattice constants in MnX alloys as a function of 
applied strain along the $a$ axis. 
The $a$ lattice parameter is linearly increased and the $b$ and $c$ lattice constants are relaxed.
In figure \ref{fig:coefficient}, 
we show the calculated values of $K_{me}$ evaluated at all values of strain 
and make a linear fit to extract the slope, 
which is used as the term $\frac{K_{me}}{\varepsilon_{bb} - \varepsilon_{aa}}$ in the Eq. (2) in the main text.
As mentioned in the main paper, the Young's moduli were taken from previous 
results.\cite{wang2013first,wang2013structural}
However, we were unable to find any values for the Young's modulus of MnIr.
To calculate the magnetostriction coefficient for MnIr, 
we adopted the bulk modulus from Materials Project \cite{materialspro} and 
used the the relation
$Y (GPa) = 3K(1-2\nu)$
where Y, K, and $\nu$ are the Young's modulus, bulk modulus, and Poisson's ratio, respectively.

\clearpage
\newpage
%